# Different Photostability of $BiVO_4$ in Near-pH-Neutral Electrolytes

*Siyuan Zhang\*, Ibbi Ahmet, Se-Ho Kim, Olga Kasian, Andrea M. Mingers, Patrick Schnell, Moritz Kölbach, Joohyun Lim, Anna Fischer, Karl J. J. Mayrhofer, Serhiy Cherevko, Baptiste Gault, Roel van de Krol\*, Christina Scheu\**

Dr. S. Zhang, S.-H. Kim, Dr. O. Kasian, A. M. Mingers, Dr. J. Lim, Dr. B. Gault, Prof. C. Scheu
Max-Planck-Institut für Eisenforschung GmbH, Max-Planck-Straße 1, 40237 Düsseldorf, Germany
E-mail: siyuan.zhang@mpie.de, c.scheu@mpie.de

Dr. I. Ahmet, P. Schnell, Dr. M. Kölbach, Prof. R. van de Krol
Institute for Solar Fuels, Helmholtz-Zentrum Berlin für Materialien und Energie GmbH, 14109 Berlin, Germany
E-mail: roel.vandekrol@helmholtz-berlin.de

Dr. O. Kasian
Helmholtz-Zentrum Berlin GmbH, Helmholtz Institut Erlangen-Nürnberg, Hahn-Meitner-Platz 1, 14109 Berlin, Germany

Prof. A. Fischer
Institute of Inorganic and Analytical Chemistry, University of Freiburg, Albertstrasse 21, 79104 Freiburg, Germany
Freiburg Center for Interactive Materials and Bioinspired Technologies, George-Köhler-Allee 105, 79110 Freiburg, Germany

Prof. K. J. J. Mayrhofer, Dr. S. Cherevko
Helmholtz Institut Erlangen-Nürnberg for Renewable Energy (IEK-11), Forschungszentrum Jülich, 91058 Erlangen, Germany

Dr. B. Gault
Department of Materials, Royal School of Mines, Imperial College London, London, SW7 2AZ, UK

Abstract: Photoelectrochemical water splitting is a promising route to produce hydrogen from solar energy. However, corrosion of semiconducting photoelectrodes remains a fundamental challenge for their practical application. The stability of $BiVO_4$, one of the best performing photoanode materials, is systematically examined here using an illuminated scanning flow cell to measure its dissolution *operando*. The dissolution rates of $BiVO_4$ under illumination depend on the electrolyte and decrease in the order: borate (pH=9.3) > phosphate (pH=7.2) >

citrate (pH=7.0). BiVO$_4$ exhibits an inherent lack of stability during the oxygen evolution reaction (OER), while hole-scavenging citrate electrolyte offers kinetic protection. The dissolution of Bi peaks at different potentials than the dissolution of V in phosphate buffer, whereas both ions dissolve simultaneously in borate buffer. The life cycle of a 90 nm BiVO$_4$ film is monitored during one hour of light-driven OER in borate buffer. The photocurrent and dissolution rates show independent trends with time, highlighting the importance to measure both quantities *operando*. Dissolution rates are correlated to the surface morphology and chemistry characterized using electron microscopy, X-ray photoelectron spectroscopy and atom probe tomography. These correlative measurements further the understanding on corrosion processes of photoelectrodes down to the nanoscopic scale to facilitate their future developments.

Photoelectrochemical (PEC) water splitting is a promising technology to harvest solar energy and store it directly as H$_2$ fuel.[1-4] A semiconducting photoelectrode functions both as a solar absorber and an electrode that drives a half-cell reaction. Many solar absorbers developed in the photovoltaic industry, *e.g.* Si, III-V, II-VI and halide perovskite compounds, have been integrated into PEC devices.[1-4] Direct contact of these materials with water, however, must be avoided to protect them from anodic corrosion and associated degradation. This motivates ongoing developments of corrosion-resistant metal oxide semiconductors for OER photoanodes.[5-8] Thus far, BiVO$_4$ has been established as one of the best performing photoanode materials,[9-14] with reported photocurrents approaching its theoretical limit of 7.5 mA cm$^{-2}$ under AM1.5G illumination.[14]

Although metal oxide semiconductors are much more stable in water than the conventional semiconductors mentioned above, few candidates have shown stability beyond a thousand hours.[15,16] Pourbaix diagrams[17] provide stability guidelines for equilibrium conditions, *i.e.*, without illumination. However, the quasi Fermi level of the holes shifts to more negative potentials under illumination, which can trigger additional anodic reactions (including the OER) and redefine the stability of the photoanode.[18-21] In practice, the photostability is often governed by the kinetics of PEC processes,[19] where recombination and the desired OER reaction compete with dissolution reactions. For $BiVO_4$, a gradual decrease in photocurrent during PEC operation has been reported.[10,22] Dissolution of stoichiometric $BiVO_4$ was first quantified in buffered phosphate electrolytes by *ex situ* measurements using inductively coupled plasma mass spectrometry (ICPMS).[23] Later studies revealed $BiVO_4$ dissolution in buffered borate[13] and citrate[24] electrolytes. However, the reports used different $BiVO_4$ photoanodes and PEC reaction conditions to study photocorrosion in various electrolytes, and observed different dissolution behaviors. In borate electrolytes, dissolution was only measured on surfaces of OER catalysts that embed the $BiVO_4$ photoanode, where stable operation for 1100[12] and 500[13] hours was demonstrated.

To close the knowledge gaps on the intrinsic photostability of $BiVO_4$, we systematically studied the dissolution behavior in all three pH-buffered electrolytes using thin (~90 nm) and compact $BiVO_4$ films prepared by pulsed laser deposition (PLD).[25] Fresh electrolyte flowed through an illuminated scanning flow cell (SFC), in which 1 mm$^2$ of $BiVO_4$ surface was exposed to the electrolyte, and then to an ICPMS that analyzes Bi and V dissolution *operando*.[24,26] As shown in **Figure 1**a, cyclic voltammetry (CV) scans on $BiVO_4$ enabled simultaneous evaluation of its activity (photocurrent density in Figure 1b), stability (dissolution curves in Figure 1c, 1d), and their dependence on the applied potential. Similar photocurrents are measured in phosphate and borate electrolytes, and both are much lower than in citrate. This is because citrate ions scavenge photo-generated holes on the $BiVO_4$ surface by the faster citrate oxidation reaction than OER. The corresponding CV scans without illumination (Figure S1) show negligible dark currents in all three electrolytes.

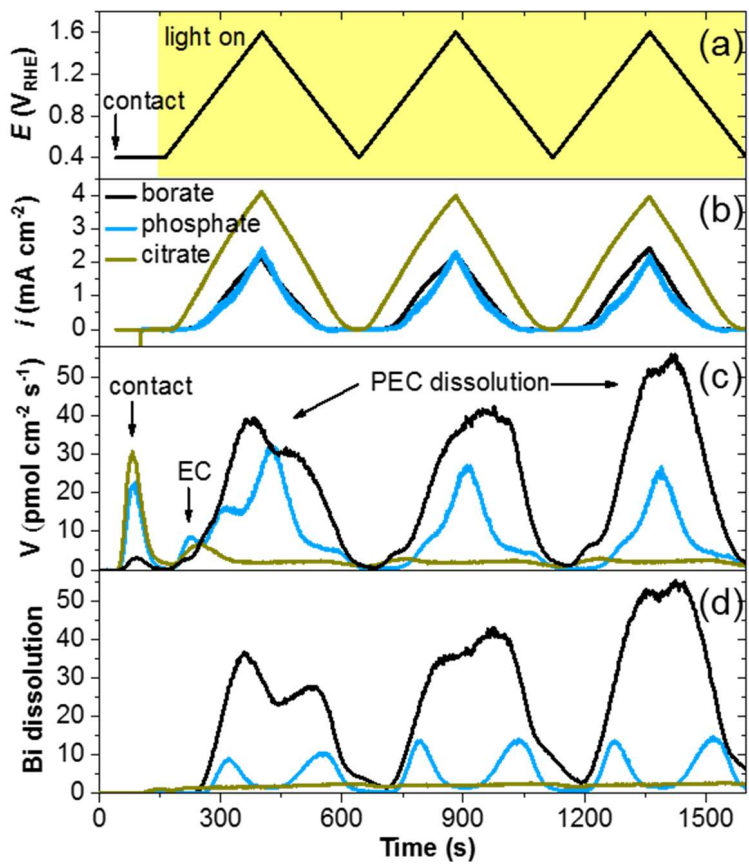

**Figure 1.** Time profiles of (a) applied potential to BiVO$_4$ (duration under illumination is highlighted in yellow) against the reversible hydrogen electrode (RHE) scale, (b) the current density $i$, (c) V and (d) Bi dissolution rates in borate (black), phosphate (blue), and citrate (dark yellow).

Different dissolution rates in three electrolytes are evident in Figure 1c, 1d. By integrating the rates, ~9, 3, and 1 nmol cm$^{-2}$ of BiVO$_4$ is dissolved during the first CV cycle in borate, phosphate, and citrate, respectively.[27] To compare, little dissolution is observed without illumination after transient V dissolution upon contact with the electrolyte (Figure S1). The same transient dissolution peaks are labeled as "contact" and "EC" (electrochemical) dissolution in Figure 1c. Bi and V dissolution after 300 s, referred to as "PEC dissolution", are only observed under illumination.

The PEC dissolution peaks repeat themselves in consecutive CV cycles, but look very different in the three electrolytes, as shown in Figure 1 and Figure S2 (replot of Figure 1 versus the potential). In borate, V and Bi dissolve simultaneously and reach their maxima at 1.6 and 1.2 V$_{RHE}$ during the respective anodic and cathodic scans. In phosphate, V dissolution peaks at ~1.6 $V_{RHE}$, whereas Bi dissolution peaks at 1.1 and 0.8 $V_{RHE}$ in the respective anodic and cathodic scans. In citrate, the dissolution rates hardly vary between 0.4 and 1.6 $V_{RHE}$. Around neutral pH, buffer ligands are the major ion species in the electrolytes with concentrations >$10^{-2}$ M, whereas [OH$^-$] lies between $10^{-7}$ (pH=7) and $10^{-4}$ M (pH=10). Therefore, buffer ligands have a deciding influence on the stability of BiVO$_4$. Hole-scavenging citrate promotes the oxidation kinetics (photocurrent onset <0.4 $V_{RHE}$) so that the slower photocorrosion reactions (*e.g.* Bi(III) oxidation) are kept at constant rates.[24] In comparison, the photocurrent onset potentials for OER in borate and phosphate electrolytes are more anodic, ~0.8 $V_{RHE}$, and distinct dissolution peaks are observed at more positive potentials. The PEC dissolution rates in borate and phosphate seem therefore kinetically coupled to the OER, a kinetically demanding reaction known for driving dissolution of many anode materials, including the more stable IrO$_2$.[28]

In phosphate solutions, Bi dissolution slows down at more positive potentials whereas V dissolution keeps increasing (Figure 1c, 1d). This suggests that the BiVO$_4$ surface becomes Bi-rich at higher anodic potentials, before Bi is further dissolved in the subsequent cathodic scan. The presence of a Bi-rich layer has been hypothesized,[10] and recently observed on a BiVO$_4$ surface immersed in phosphate by ambient pressure hard X-ray photoelectron spectroscopy, evidencing BiPO$_4$ formation under illumination and its dissolution in the dark.[29]

In borate solutions, dissolution was only reported *ex situ* on BiVO$_4$ covered by FeOOH/NiOOH catalysts, exemplifying enhanced OER kinetics and lower dissolution of primarily V.[13] In contrast, we find that the bare BiVO$_4$ surface has poor stability during OER in borate, showing the highest dissolution rates among the three studied electrolytes. Furthermore, as shown in Figure 1c,1d, the dissolution rates in borate keep increasing from the first to the third CV cycle.

The changing dissolution rates are also monitored in an accelerated degradation test shown in **Figure 2**, as most of the examined BiVO$_4$ photoanode dissolved within one hour of OER at 1.6 V$_{RHE}$. Throughout the life cycle, the dissolution rates first increased to ~60 pmol cm$^{-2}$ s$^{-1}$,[27] then decreased with time toward zero. The amount of dissolved material is determined by integration, which enables a reconstruction of the film thickness with time (right vertical axis in Figure 2a). A drop in the photocurrent during the first illumination cycle was reversed after the break, during which the SFC was detached from the BiVO$_4$ surface and then approached back to remove evolved O$_2$ bubbles from the surface. Such breaks were introduced three more times during the chronoamperometric measurement.

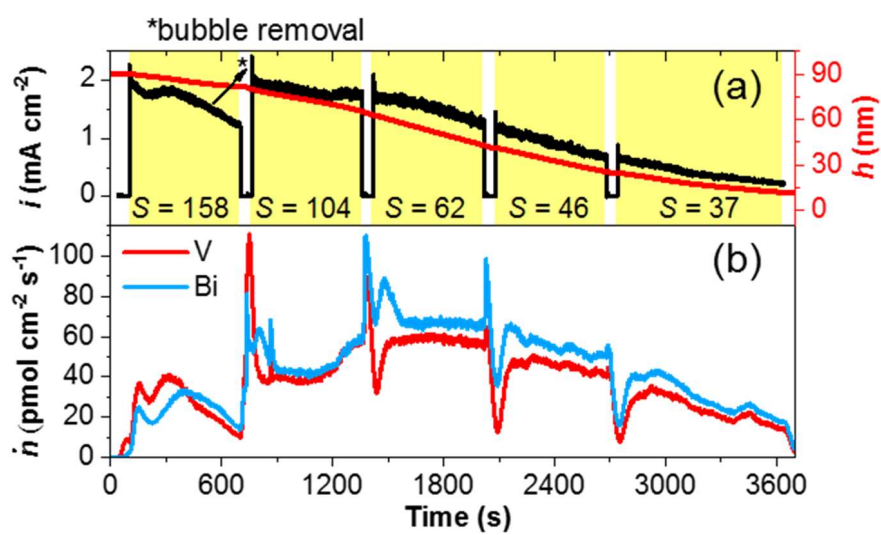

**Figure 2.** Time profiles of (a) photocurrent density $i$ (black), film thickness $h$ (red), (b) dissolution rates $\dot{n}$ of BiVO$_4$ held at 1.6 V$_{RHE}$ in borate. The duration under illumination are highlighted in yellow, with calculated stability number $S$ shown in each illumination cycle.

Varying dissolution rates can be correlated to chemical composition and morphology of BiVO$_4$ photoanodes, characterized before and after OER. X-ray photoelectron spectroscopy (XPS) reveals that as-synthesized BiVO$_4$ films have a V-rich surface (59±1%, Table S1), which correlates to the transient dissolution of primarily V amounting to ~2 nmol cm$^{-2}$ (Figure 1, S1).[27] After ~20 min OER, the surface becomes close to stoichiometry (51±2% V, Table S1), on which stoichiometric PEC dissolution proceeded in borate.

As shown in **Figure 3**a, as-synthesized BiVO$_4$ films are composed of crystallites with lateral sizes of 200−500 nm. The surface becomes porous after OER in borate: The arrow in Figure 3f points to a pore, while cross-section images (Figure 3b, 3g) reveal porous features throughout the thickness of the film. As photocorrosion progresses, the once embedded pores emerge to the surface (Figure 3g), while the BiVO$_4$ film remains as the monoclinic scheelite phase (Figure 3h, 3i). The porous morphology causes the effective surface area to increase, which explains the initial increase in the dissolution rates. Dissolution stops at the end of the one-hour measurement, and Figure S4 shows that little BiVO$_4$ remains within the exposed area.

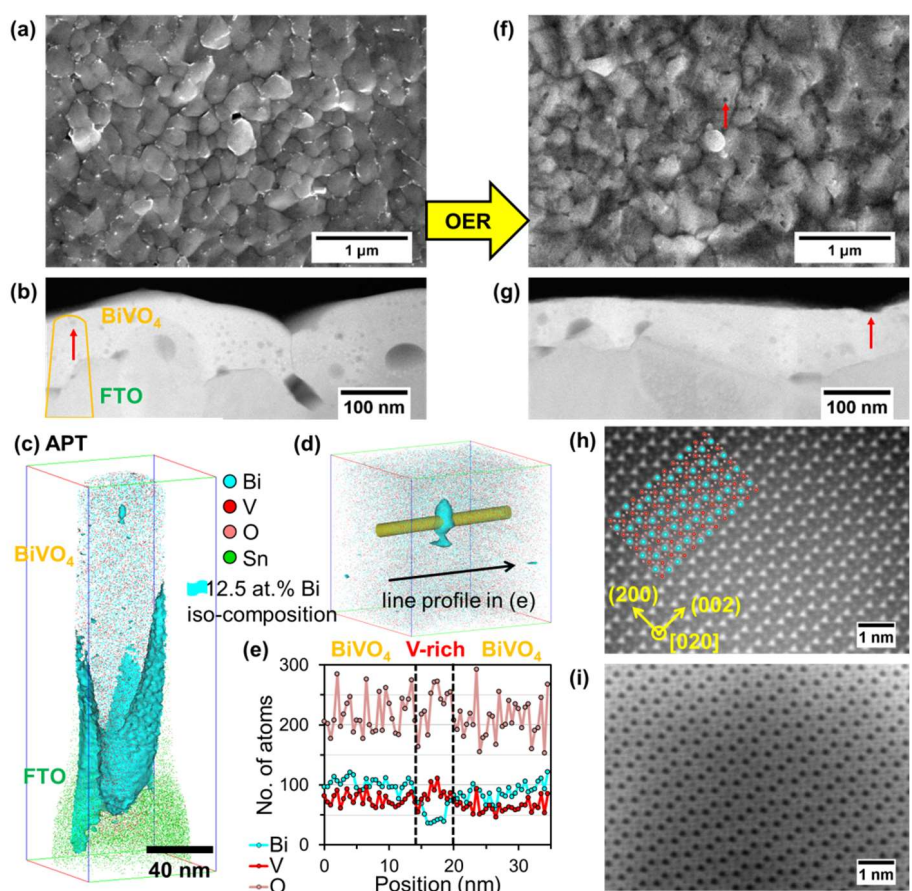

**Figure 3.** Morphology and chemical composition of BiVO$_4$ films (a-e) before and (f-i) after 20 min OER in borate. (a,f) Top view scanning electron microscopy (SEM), (b,g) cross-section view scanning transmission electron microscopy (STEM), atomically resolved (h) high angle annular dark field (HAADF), (i) annular bright field (ABF)-STEM micrographs with overlaid monoclinic scheelite BiVO$_4$ structure, (c) 3D atom map of a needle-shaped APT sample from a similar region as the yellow outline in (b), (d) sectioned volume with a cylindrical $\varPhi$4.3×35 nm$^3$ region across a feature enveloped by the iso-composition surface of 12.5 at.% Bi, and (e) detected numbers of Bi, V, O atoms within the cylinder.

Larger pores in the cross-section images (Figure 3b, 3g) are present along the interface between BiVO$_4$ and fluorine-doped SnO$_2$ (FTO) substrate. As the FTO surface is rough, limited surface diffusion of the deposited species during room temperature PLD growth may explain the incomplete coverage. We also observe smaller (~10 nm), round-shaped dark features, whose composition was further examined by atom probe tomography (APT). One of these spherical features is highlighted in Figure 3d, and shown to contain fewer Bi atoms (Figure 3e), indicating a V-rich oxide. Once PEC dissolution exposes such V-rich oxides to the surface, they are unstable in aqueous electrolytes[24] and their preferential dissolution leads to a porous surface.

Depletion in photocurrents is conventionally used as the indicator for photocorrosion.[10,13,22-23] For photoanodes, both the OER and the anodic corrosion reaction are driven by photo-generated holes. However, as shown in Figure 2, the steady state photocurrents varies only little between the first two illumination cycles, even though by then one third of the film had dissolved (90 to 60 nm). This is because BiVO$_4$ has a carrier diffusion length of ~70 nm,[30,31] so that only holes generated in the first 70 nm of the film can contribute to the photocurrent. As a result, the photocurrent of thicker films is not affected by photocorrosion. Below the observed transition at ~60 nm (from the third illumination cycle in Figure 2), which is indeed close to the carrier diffusion length of BiVO$_4$, the surface hole population becomes limited by the absorber volume, leading to decreasing photocurrents and dissolution rates.

We further evaluate the relative stability of BiVO$_4$ with respect to the OER, using the stability number, $S$, a quantity introduced for Ir-based OER catalysts as the ratio between numbers of evolved O$_2$ molecules and dissolved Ir atoms in the electrolyte.[28] To determine $S$ for BiVO$_4$, we use the average value of dissolved Bi and V atoms in the denominator, since both elements dissolve at an approximately equal rate in borate. As shown in Figure 2, the relative stability drops monotonically from $S$=158 averaged over the first illumination cycle to $S$=37 during the fifth. In comparison, WO$_3$ photoanodes studied in sulfuric acid have similar $S$ numbers, ranging from 100 to 200,[26] whereas OER anodes of commercial polymer electrolyte membrane electrolyzers made of iridium compounds are remarkably more stable, with $S=10^4–10^7$.[28] Having $S$ well over unity indicates that the OER is the main anodic reaction, which could kinetically inhibit photocorrosion.[19] However, with $S$ on the order of $10^2$, bare BiVO$_4$ surfaces only demonstrate ~1 h lifetime in the borate buffer. The lifetime of an electrode is proportional to $S$ and the loading (film thickness).[28] Since BiVO$_4$ photoanodes must remain thin (~100 nm) in order to ensure efficient carrier collection,[30,31] they need to have $S$ numbers on the order of $10^6$ in order to reach the >$10^4$ h lifetime required for practical applications. We demonstrate that kinetic inhibition by citrate oxidation only slows down the dissolution by one order of magnitude. Therefore, passivating layers including OER catalysts are indispensable to protect BiVO$_4$ surface from photocorrosion, having thus far demonstrated stable operation for 1100[12] and 500[13] hours.

In summary, using illuminated SFC coupled to ICPMS, we observed high dissolution rates of BiVO$_4$ during light-driven OER in borate and phosphate electrolytes. In phosphate electrolyte, Bi and V dissolve at different potential ranges, whereas their dissolution in borate is synchronous. Photocurrents alone cannot reflect the progression of photocorrosion, especially for thicker photoabsorbers with limited carrier diffusion lengths. Therefore, we propose to rely on *operando* dissolution measurements to evaluate photostability. Evolution of PEC dissolution rates can be correlated to the surface and microstructure of BiVO$_4$. Nanoscopic V-rich oxide regions are identified by APT and found dispersed within the film. As photocorrosion progresses, exposure of these V-rich oxides to the electrolyte leads to their preferential dissolution and formation of pores. The toolset to characterize both sides of the semiconductor-liquid junction can facilitate future investigation on protection strategies against photocorrosion.

**Experimental Section**

*Photoanode preparation*: BiVO$_4$ was deposited on FTO coated glass (TEC7, Pilkington) by PLD in vacuum (<2×10$^{-6}$ mbar) at room temperature. The laser fluence, spot size and pulse frequency were set to 1.5 J cm$^{-2}$, 1.3×2 mm$^2$ and 10 Hz, respectively. The target-to-substrate distance was 60 mm. Further details on the PLD system and the BiVO$_4$ target are reported elsewhere.[25] After deposition, the films were annealed at 450 °C in air for 2 h.

*Illuminated SFC-ICPMS setup*: As detailed in another report,[24] PEC experiments were performed through a SFC controlled by a Reference600 potentiostat (Gamry). The SFC was positioned to contact the BiVO$_4$ working electrode and limits the area to 1 mm$^2$. A Pt-wire (0.5 mm, 99.997%, Alfa Aesar) counter electrode and a saturated Ag/AgCl reference electrode (Metrohm) were respectively placed in the inlet and outlet channels of the SFC. The borate buffer (pH=9.3–9.4) was prepared from Na$_2$B$_4$O$_7$ (15 mM, Merck, suprapure). The phosphate buffer (pH=7.1–7.2) was prepared from KH$_2$PO$_4$ (5.2 mM, Merck, suprapure) and Na$_2$HPO$_4$ (8.2 mM, Merck, suprapure). The citrate buffer (pH=6.9–7) was prepared from C$_6$H$_8$O$_7$·H$_2$O (15 mM, Merck, p.A.) and NaOH (44.1 mM, Merck, Titrisol). To ensure accurate evaluation of V$_{RHE}$, electrolytes were measured by a MultiLab540 pH-meter (WTW) on a daily basis. Electrolytes were pumped through the SFC at 3.4 µL s$^{-1}$ and analyzed online in a NexION300X ICPMS (Perkin Elmer). A Superlite S04 light (Lumatec) filtered to 400–700 nm was guided to illuminate the front side of BiVO$_4$, with intensity calibrated to 100 mW cm$^{-2}$ using a Si-diode photometer (Newport).

*Structural characterization*: XPS was measured on a Quantera II (Physical Electronics) using a monochromatic Al-Kα X-ray source (1486.6 eV) operated at 15 kV and 25 W, and analyzed by the Casa XPS software. SEM micrographs were taken using a Scios microscope (Thermo Fisher) operated at 10 kV using a secondary electron detector. STEM was performed on a Titan Themis microscope (Thermo Fisher) operated at 300 kV, with an aberration-corrected electron probe of 23.8 mrad convergence. HAADF and ABF micrographs were formed using electrons scattered to 73−200 and 8−16 mrad, respectively. APT measurements were conducted on a LEAP5000XS (CAMECA) at 70 K base temperature, in pulsed laser mode at 2% detection rate, 60 pJ pulse energy, and 125 kHz pulse frequency. 3D reconstruction and composition were analyzed using IVAS3.8.4 software.